# Vehicle-to-Grid Integration: Ensuring Grid Stability, Strengthening Cybersecurity, and Advancing Energy Market Dynamics

BILAL AHMAD[a], JIANGUO DING[b], TAYYAB ALI[c], DOREEN SEBASTAIN SARWATT[a], RAMSHA ARSHAD[a], ADAMU GASTON PHILIPO[a], HUANSHENG NING[a]

[a]*School of Computer and Communication Engineering University of Science and Technology Beijing China*
[b]*Department of Computer Science Blekinge Institute of Technology 371 79 Karlskrona Sweden*
[c]*Department of Electrical and Computer Engineering International Islamic University (IIU) Islamabad Pakistan*

---

## Abstract

The increasing adoption of electric vehicles has spurred significant interest in Vehicle-to-Grid technology as a transformative approach to modern energy systems. This paper presents a systematic review of V2G systems, focusing on their integration challenges and potential solutions. First, the current state of V2G development is examined, highlighting its growing importance in mitigating peak demand, enhancing voltage and frequency regulation, and reinforcing grid resilience. The study underscores the pivotal role of artificial intelligence and machine learning in optimizing energy management, load forecasting, and real-time grid control. A critical analysis of cybersecurity risks reveals heightened vulnerabilities stemming from V2G's dependence on interconnected networks and real-time data exchange, prompting an exploration of advanced mitigation strategies, including federated learning, blockchain, and quantum-resistant cryptography.Furthermore, the paper reviews economic and market aspects, including business models (V2G as an aggregator or due to self-consumption), regulation (as flexibility service provider) and factors influencing user acceptance shaping V2G adoption. Data from global case studies and pilot programs offer a snapshot of how the rule has been put into place, at different paces across regions. Finally, a multi-layered framework that incorporates grid stability resilience, cybersecurity resiliency, and energy market dynamics is suggested by the study provides strategic recommendations to enable scalable secure and economically viable V2G deployment.

---

*School of Computer and Communication Engineering, University of Science and Technology Beijing, China
**E-mail: ninghuansheng@ustb.edu.cn (HUANSHENG NING)

---

## 1. Introduction

The electric vehicles (EVs) Transition has progressed rapidly in the last year with more emphasis on reducing dependence of fossil fuels and general concern for environment. Globally, the number of EVs on the road has surged to over 5.1 million by 2018 — largely in China (45%), followed by Europe (24%) and United States (22%) [1]. V2G has the ability to provide electricity back to the power grid and permits the energy flow in both directions between EVs and a The technology allows EVs to feed back power into the grid when required, helping balance performance with the capacity to deal with renewable energy intermittency [2]. EVs can charge when electricity is cheap and discharge when costly, they are the ultimate temporal and spatial energy arbitrage tools that can help manage transmission constraints and improve system efficiency [3, 4].

V2G is needed to stabilize the grid: load smoothing, (more) renewable penetration, economic efficiency and network losses[5]. Up to 40% less grid load with smart charging and energy storage enabling renewable integration into distribution [6, 7, 8]. Optimized strategies like APD and smart grid concepts ensure efficient energy scheduling, minimize losses, and foster a resilient energy ecosystem [9, 10, 11]. V2G enables EV owners to earn revenue through demand response and ancillary services while optimizing grid efficiency with dynamic pricing and real-time energy trading [2, 12, 13]. This arrangement creates revenue opportunities for owners while supporting grid stability [14, 15, 1].

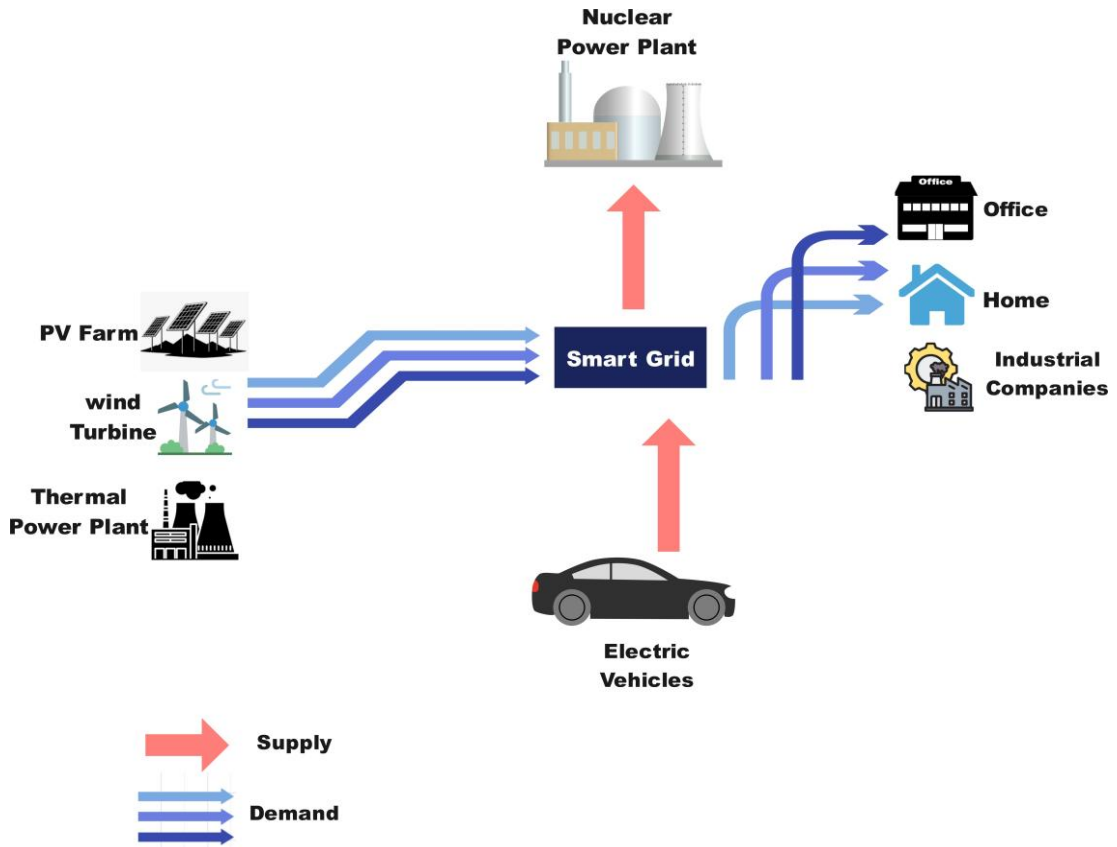

Figure 1: V2G integration and renewable sources

However, V2G systems face cybersecurity vulnerabilities due to reliance on communication and smart grid infrastructure, particularly risks of cyber-physical attacks. Strong cybersecurity measures, including advanced encryption and secure protocols, are essential to prevent data breaches [2, 3, 4]. Challenges to implementation include renewable integration, regulatory barriers, grid complexities and the need for predictable load for dispatchability [12, 13, 14]. Faced with such economic and policy challenges, strategic stakeholder engagement and innovative business models are required [15, 16, 17].

Transactions of energy are efficiently and securely performed through blockchain technology and VPPs [18]. Approaches such as cooperative game theory, predictive algorithms and hybrid optimization methods provide insight on an efficient V2G operation [16, 17, 19]. Turning to the performance of EV charging sta-tions placement and solutions like On-board Charging is important in bringing changes to the adoption [12, 13, 14]. To drive this success, the V2G system needs to solve some key challenges that would boost grid performance and increase its stability and reliability. By relieving pressure on infrastructure and lowering carbon gas emissions this technology support sustainable energy by paving the way for faster decarbonization [9, 10, 11]. V2G still holds promise as a part of the solution to meet sustainability targets, but there are challenges in scaling V2G due to its dependence on initial stakeholder buy-in around technology development [6, 7, 8]. The implementation of V2G systems to be successful requires overcoming these issues as they directly interfere with the primary objective of enhancing grid efficiency, stability and reliability. In this research, a solution for the integration of V2G was presented and along with it an appropriate definite thing to do for cybersecurity that would keep these critical control systems protected. It provides cutting-edge grid stability solutions and underscores the market need that will help to drive widespread V2G adoption, spawning strong infrastructure and financial incentives within an energy ecosystem demonstrating strength and agility. The main contribution of this article can be summarized as follows.

1. This article identifies and analyzes the key challenges associated with maintaining grid efficiency, stability, and reliability in V2G integration. It highlights the critical role of AI and ML in optimizing energy management strategies, enhancing load forecasting accuracy, and enabling real-time grid control to mitigate operational uncertainties.

2. The study examines the growing cybersecurity threats inherent in V2G systems, particularly vulnerabilities in communication protocols and control mechanisms that could compromise grid stability. Emphasis is placed on advanced protective solutions including federated learning, blockchain-based authentication, and post-quantum cryptographic techniques to ensure the resilience and integrity of critical grid infrastructure.

3. The article evaluates the market viability of V2G systems by investigating dynamic pricing models, regulatory policies, and incentive structures. These mechanisms are shown to enhance economic feasibility while encouraging broader stakeholder participation, ultimately contributing to improved grid performance and sustainable energy transition.

This study adopts an eight-section structure to systematically examine Vehicle-to-Grid (V2G) integration. Section 1 introduces electric vehicles (EVs) and V2G integration while outlining research objectives. Section 2 analyzes grid stability and reliability challenges, followed by Section 3's investigation of cybersecurity risks and mitigation strategies. Section 4 assesses energy market dynamics and emerging business models, while Section 5 explores their interdependencies with grid stability and cybersecurity. Regulatory and implementation challenges are addressed in Section 6, followed by future research recommendations in Section 7. The paper concludes with key insights in Section 8. Logical transitions between sections ensure coherence and readability.

### 1.1. Procedure to Conduct Review

The systematic review methodology encompassed four key phases, beginning with database exploration across IEEE Xplore„ACM, SpringerLink, ScienceDirect, and MDPI followed by keyword selection combining "Vehicle-to-Grid (V2G)" with concepts such as "Electric Vehicles (EVs)," "Cybersecurity," "Charging Infrastructure," "Grid Stability," "Peak Load Management," "Power Management," "Voltage Stability," and "Cyber-Physical Attack (CPS)." This was supplemented by examining references cited in identified review papers, and after removing duplicates and irrelevant records, 69 publications were evaluated against specific inclusion criteria focusing on V2G integration aspects, ultimately yielding 20 relevant papers whose temporal distribution showed an upward publication trend between 2020 and 2025 across various digital libraries, Fig.2 illustrate the publication trend, respectively.

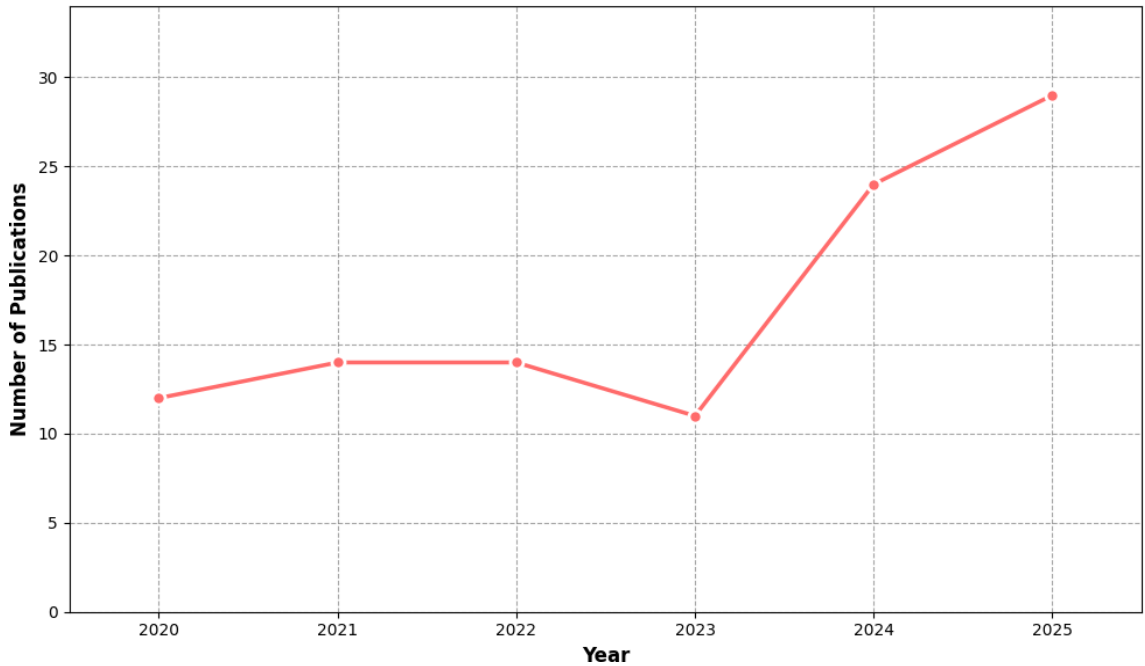

Figure 2: Publication Trend Analysis (2020-2025)

*1.2. Related Work*

The integration of electric vehicles into modern power systems has sparked extensive academic and industrial research, particularly on the consequences for grid stability, cybersecurity, and market dynamics.

**Grid Stability, Efficiency, and Control:** Extensive research has explored the impact of EVs on power grid performance, identifying key challenges such as grid overloading, voltage instability, and reduced system reliability, while also acknowledging the potential of EVs to contribute to load shifting and ancillary grid support services [20]. However, much of the existing literature overlooks the complex technical demands of full-scale V2G integration, particularly in areas such as frequency regulation, voltage control, harmonics mitigation, and infrastructure interoperability. These challenges necessitate the adoption of advanced control strategies and AI-enabled security frameworks [20]. To address these issues, AI and ML have emerged as powerful tools, offering enhanced capabilities in demand forecasting, energy scheduling, and real-time grid control. AI is playing an increasingly critical role in improving energy efficiency, operational transparency, and the integration of renewable energy sources. Never-theless, several international studies emphasize the need for greater investment in AI-driven energy system modelling and implementation [21]. Centralized coordination approaches have demonstrated the potential to mitigate grid stress during periods of high EV penetration by optimizing operational costs and reducing emissions. However, these strategies often require substantial infrastructure investments and raise concerns regarding battery degradation and lifecycle management[22]. In contrast, decentralized and hierarchical distributed charging architectures aim to optimize charging processes through various models and algorithms that have been proposed to improve system scalability and reduce communication overhead [23, 24]. Additionally, the integration of EVs within microgrids, especially when combined with distributed energy resources, has shown measurable improvements in system reliability, operational flexibility, and coordinated energy storage utilization [25].

**Cybersecurity Challenges in V2G Integration:** Vehicle-to-Grid systems increasingly depend on real-time data exchange and interconnected communication protocols, rendering them highly susceptible to cybersecurity threats. Existing research highlights the critical need to safeguard control mechanisms, ensure data integrity, and secure communication channels against potential cyberattacks [20]. Emerging cybersecurity solutions such as federated learning, blockchain-based authentication, and post-quantum cryptographic techniques offer promising avenues to enhance system resilience, data privacy, and operational trustworthiness. However, standardized security frameworks remain underdeveloped, and real-world validation, particularly for V2V communication and distributed control systems, is still limited [26]. Ensuring the robust protection of critical V2G infrastructure requires interdisciplinary strategies that integrate advanced cryptographic methods with resilient system architectures and adaptive threat detection mechanisms [27].

**Economic Feasibility and Market Adoption:** The widespread adoption of V2G systems is closely tied to market mechanisms, consumer behaviour, and the presence of supportive regulatory frameworks. Numerous studies have explored dynamic pricing models, policy incentives, and business strategies designed to enhance the economic viability of V2G technology [28, 29]. A global review of 131 V2G pilot projects across 27 countries found that, although the majority focus on technical feasibility and economic potential, only 27% explicitly address user acceptance, revealing a significant gap in understanding the behavioural and socio-economic dimensions that influence large-scale implementation [30]. Key barriers such as concerns over battery degradation, limited financial returns, and a general lack of consumer trust remain critical obstacles. Case studies from markets like California and Denmark further highlight the limitations of current incentive structures and insufficient public awareness. Moreover, without adequate charging infrastructure, well-defined regulatory support, and targeted public policies, the scalability of V2G systems is likely to remain constrained [20, 31]. Much of the existing research continues to rely heavily on simulations, with limited focus on large-scale demonstration projects, thereby restricting insights into real-world charging behaviour, user interaction patterns, and operational integration with the grid. Addressing these research and implementation gaps is essential to fully assess battery lifecycle impacts, improve user engagement strategies, and unlock the full potential of V2G technology for grid reliability and renewable energy integration [32, 27]. Table 1 provides a comparison with earlier research, while our analysis covers all the key areas shown in the table, including grid stability, renewable energy integration, and cybersecurity concerns.

Table 1: A comparison with prior studies

| Area Explored | [26] | [33] | [34] | [35] | [36] | [37] | [38] | [27] | [39] | Our survey |
|---|---|---|---|---|---|---|---|---|---|---|
| Grid Stability | ✓ | × | × | ✓ | × | ✓ | ✓ | × | × | ✓ |
| Grid Reliability | × | × | × | ✓ | ✓ | ✓ | ✓ | ✓ | ✓ | ✓ |
| Power Management and DR | × | ✓ | × | × | ✓ | ✓ | ✓ | × | ✓ | ✓ |
| Renewable Energy Integration | × | × | × | ✓ | × | × | × | × | ✓ | ✓ |
| Infrastructure Development | ✓ | ✓ | ✓ | ✓ | ✓ | × | × | ✓ | ✓ | ✓ |
| Battery Degradation Concerns | × | × | ✓ | × | × | × | × | ✓ | × | ✓ |
| Cybersecurity | ✓ | ✓ | × | × | × | × | × | × | × | ✓ |
| Data Privacy and Security | ✓ | ✓ | ✓ | × | × | × | × | × | × | ✓ |
| Market Dynamics | ✓ | × | ✓ | ✓ | × | ✓ | × | ✓ | × | ✓ |
| Interoperability Issues | × | ✓ | ✓ | ✓ | ✓ | ✓ | ✓ | ✓ | ✓ | ✓ |

## 2. Grid Stability, Reliability, and Efficiency in V2G Integration

Vehicle-to-Grid systems exert a substantial influence on the electrical grid, accounting mechanisms, and overall reliability. Ensuring network stability is crucial for the dependable and efficient operation of V2G services as shown in Fig.3. Instability can induce voltage fluctuations, heightened peak demands, and harmonic distortions, thereby undermining network reliability. Capacity limitations determine the feasibility of scaling V2G implementations and may pose significant challenges if the network cannot accommodate additional loads. Dynamic pricing models related to V2G may lead to sustainability concerns, contributing to voltage instability and increased peak demands, which further affect overall system performance. Table 2 highlights the essential role of V2G in enhancing grid stability, optimizing charging infrastructure, and integrating renewable energy sources. Implementing distributed charging infrastructure can mitigate unpredictability and bolster the network's capacity to manage diverse energy demands.

### 2.1. Overview of Grid Stability Challenges with V2G

V2G integration enhances power system reliability through peak and valley regulation, frequency regulation, voltage regulation, and renewable energy consumption, yet presents several technical challenges. These include battery degradation due to increased cycling, compatibility issues and high costs associated with bidirectional charging equipment, and the precise synchronization requirements with grid voltage, frequency, and phase parameters to maintain network stability [40]. Furthermore, user concerns regarding battery longevity and the need for standardized technical compatibility across various EV models and chargers are critical barriers to widespread adoption.

Effective DR management is equally important for optimizing load profiles, with RES and EV integration playing pivotal roles in supporting DR initiatives for utilities and consumers [41]. Real-time data exchange enabled through sophisticated communication technologies permits improved load forecasting and energy distribution, which enhances overall grid efficiency [36]. However, challenges such as battery degradation, substantial infrastructure costs, and social barriers remain in V2G integration. Optimization techniques such as GA and PSO can be employed to address these issues [42].

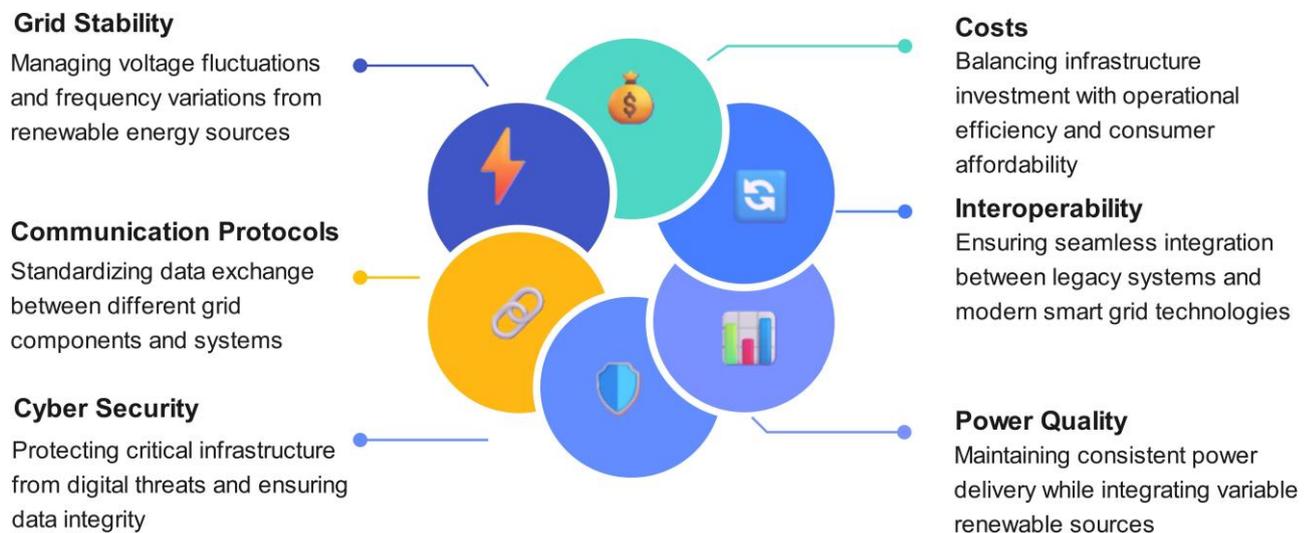

Figure 3: Grid Integration Challenges

BESS are instrumental in stabilizing the grid through power smoothing, voltage regulation, and frequency support, particularly as more EVs and RES are integrated [39]. Vehicle-to-Grid technology uses EVs to enhance grid stability, reducing power losses and improving voltage stability through metrics like min-imum voltage and Voltage Stability Index. However, EV charging stations cause power losses and congestion in radial systems due to erratic charging [38]. Advanced control strategies manage active power, battery State-of-Charge, and frequency regulation for grid reliability [43]. Scalability for large EV fleets and widespread V2G adoption remains a challenge [44].

V2G aligns charging and discharging with energy needs and user preferences. Robust scheduling, AI-driven optimization, and standardized protocols are essential. Challenges include computational demands, real-time communication, cybersecurity, and EV battery health [45]. Bidirectional converters are essential for regulating power flow and facilitating the integration of renewable energy sources, such as solar power, thereby enabling efficient energy exchange between electric vehicles and the grid. Advanced methodologies, including power factor correction, real-time load management, and intelligent scheduling supported by tools such as the Voltage Stability Index and Battery Energy Storage Systems are vital for maintaining grid stability and optimizing demand response.

Table 2: Key Aspects of V2G Integration

| Feature | Importance in V2G | Challenges | Ref |
|---|---|---|---|
| Grid Stability Support | Provides ancillary services like frequency regulation, peak shaving, and load balancing, improving grid reliability. | Dynamic load fluctuations and intermittent renewable generation complicate real-time control. | [46] |
| Scalable and efficient control for EV charging | To ensure grid stability with growing EV adoption, it lowers costs by scheduling charging during off-peak times, enhances convenience, and promotes sustainability through renewable integration and reduced emissions | Scalability limited by computational bottlenecks requiring robust distributed architectures. EV arrival/departure uncertainties and limited EVSE/parking spaces complicate real-time scheduling and forecasting capabilities | [24] |
| Communication and Coordination | Facilitates data exchange between primary and secondary converters, ensuring seamless operation and power modulation. | Signal processing delays can limit coordination, necessitating robust communication channels for effective control. | [47] |
| Smart Charging and Discharging | Utilizes advanced algorithms to manage charging and discharging based on grid needs and user preferences, optimizing energy flow and supporting solar energy integration. | Developing algorithms that balance grid demands, user needs, and battery longevity while handling variable solar energy inputs is challenging. | [48] |

Table 2: Key Aspects of V2G Integration

| Feature | Importance in V2G | Challenges | Ref |
|---|---|---|---|
| Integration with Renewable Energy | EVs can store excess solar/wind energy and discharge it when renewables are unavailable, enhancing grid stability. | Limited overlap between EV availability (e.g., overnight) and solar generation peaks (daytime) reduces synergy. | [49] |
| Advanced Control Strategies for V2G Operations | Advanced V2G control strategies optimize energy flow between EVs and the grid, enhancing stability, renewable integration, and efficiency while reducing peak loads | V2G control strategies face challenges from complex fleet coordination, battery degradation, cybersecurity risks, and inconsistent grid codes | [50] |
| Congestion Relief through V2G-Enabled BEVs | V2G-enabled BEVs reduce grid congestion by optimizing energy flow and supporting renewable integration through strategic charging and discharging | Frequent V2G cycles degrade BEV batteries, while high costs, complex coordination, and regulatory barriers limit participation in congestion management | [51] |
| Resilience to Disruptions | V2G enables EVs to supply power during disruptions, enhancing microgrid resilience through frequency regulation, peak shaving, and DER integration | V2G causes battery degradation and requires complex coordination, with variable EV availability complicating scheduling and integration | [25] |

## 2.2. Strategies for Managing Grid Stability

Vehicle-to-Grid technology allows EVs to supply power during peak demand, providing grid services while managing voltage fluctuations and harmonic distortions. Optimized charging algorithms, like GA, reduce peak loads and balance grid performance, costs, and battery life [40]. EV integration with renewable energy mitigates intermittency using EVs as storage and DTR for transmission efficiency [36]. Pilot programs show cost savings, reliability, and lower emissions, despite pricing and battery degradation challenges.

Integrating RDGs, DSVCs, and V2G-enabled EVCS enhances grid stability, reduces losses by 25% using the Spotted Hyena Optimizer Algorithm, and cuts $CO_2$ emissions [39]. Localized microgrids with these technologies reduce Energy Not Delivered by 32% during outages [43]. Real-time adjustments optimize RDG output and EV charging for balanced supply-demand and grid resilience. As renewable energy sources and EVs increasingly penetrate the market, strategic load profile optimization becomes essential for maintaining effective Demand Response capabilities [38].

Demand Side Management techniques significantly contribute to peak load reduction and cost minimization while concurrently decreasing emissions through the intelligent integration of storage technologies with renewable resources. Modern machine learning techniques can significantly increase system balance while lowering operating costs, especially SVR for load prediction and the DA for scheduling optimization [42]. V2G integration enables EVs to stabilize the grid by dynamically supplying or absorbing energy based on demand [44]. A coordinated algorithm optimizes EV charging and discharging, minimizing frequency fluctuations while respecting battery limits.

## 2.3. Enhancing Grid Reliability with V2G

Vehicle-to-Grid systems enable electric vehicles to function as both energy consumers and suppliers, facilitating bidirectional energy flow to support grid stability and renewable integration. However, their implementation faces several challenges as depicted in Figure 3. Unidirectional or grid-to-vehicle charging allows EVs to draw power from the grid as passive loads, risking grid strain during peak demand [52]. Bidirectional charging enables EVs to return power to the grid, acting as distributed energy resources. V2G enhances grid stability, mitigates renewable intermittency, and supports peak shaving through advanced bidirectional switches. The integration of renewable energy sources, like solar and wind, is supported by EV batteries' capacity to store energy and release it back into the grid [36]. This adaptability smooths out variations in the energy supply and improves grid stability overall, reducing the intermittent nature of renewable energy sources. Vehicle-to-Grid technology boosts grid reliability by using EVs to supply or absorb power, stabilizing frequency [43]. Optimized EVCS, RDGs, and DSVCs reduce power losses and voltage deviations. By utilizing EV batteries as storage solutions, V2G systems contribute to energy storage capacity within the grid [42]. This stored energy can be used during periods of high demand or low generation from renewable energy sources, enhancing the grid's ability to respond to varying energy needs. V2G enables EVs to stabilize the grid by supplying or absorbing power based on frequency [44]. A control algorithm manages charging and discharging, ensuring safe battery SOC. Vector control optimizes power flow via a DC/AC converter. The scalable system regulates frequency deviations within seconds. V2G systems facilitate EVs functioning as energy suppliers, thereby contributing to grid stabilization through the integration of renewable energy sources and load management utilizing real-time data.

### 2.4. AI and Machine Learning for Grid Management

The integration of renewable energy sources and EVs into the power grid presents challenges such as increased peak demand and voltage instability, which can impact grid reliability and performance. Advanced machine learning methodologies can effectively address these issues such as Support Vector Regression, energy demand can be forecasted with notable precision, while the Dragonfly Algorithm optimizes scheduling procedures [42]. These innovative technologies collaboratively work to improve load distribution, reduce operational costs, and foster a more stable and dependable energy infrastructure, as illustrated in Fig. 4 [39].

Artificial intelligence and machine learning algorithms enhance V2G systems by optimizing bidirectional energy transfer, utilizing predictive models to manage load balancing and peak demand with real-time data from smart grid infrastructure and aggregators [53]. Techniques such as reinforcement learning and neural networks are employed to predict energy demand and the state-of-charge of electric vehicle batteries, thereby improving grid reliability and the integration of renewable energy. Nonetheless, challenges including battery degradation and the lack of standardized protocols necessitate the development of robust AI algorithms and infrastructural improvements.

AI-driven V2G systems for electric school buses utilize machine learning and deep learning techniques to forecast energy demand and coordinate charging schedules [54]. These advancements contribute to enhanced grid stability and cost efficiency. Techniques such as agile optimization enable real-time adjust-ments to grid conditions, supporting the integration of renewable energy sources. The use of machine learning and socio-anthropological data for effective energy scheduling enhances grid support and incentivizes participation of EVs users [55]. Multi-objective optimization aims to minimize battery degradation and grid fluctuations, while social insights address user concerns such as range anxiety.

V2G plays a vital role in optimizing power demand and ensuring smart grid sustainability, addressing issues such as dynamic pricing, EV charging uncertainties, battery degradation, power quality, and charger standardization [56]. AI solutions, including genetic algorithms and hybrid PSO-ACO methods, alongside advanced control strategies and policy frameworks, are employed for efficient V2G implementation. The integration of V2G systems with renewable energy sources requires advanced methodologies to mitigate grid instability and peak demand challenges. Incorporating sophisticated machine learning techniques like Support Vector Regression and the Dragonfly Algorithm can optimize energy demand prediction and scheduling, thereby enhancing load distribution.

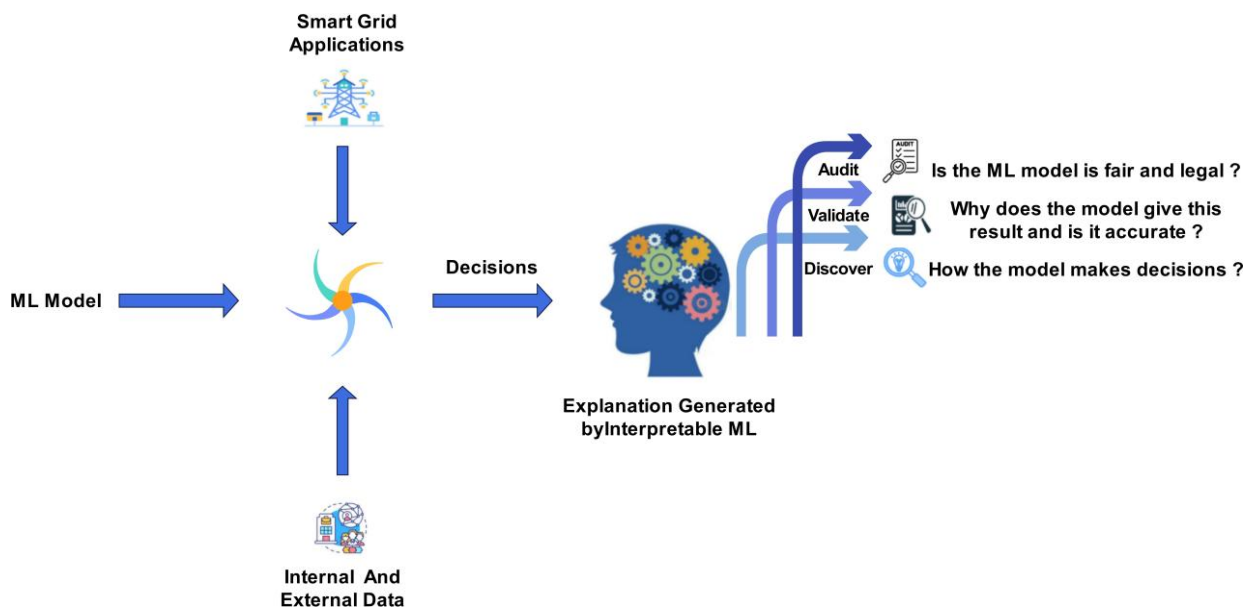

Figure 4: An illustration of AI-powered Grid Management

## 3. Cybersecurity Challenges and Solutions in V2G Systems

Cybersecurity in V2G systems faces challenges from privacy, insecure communication, and data manipulation vulnerabilities (as shown in Fig. 5), enabling attacks like privacy violations and man-in-the-middle attacks. Despite research on protective measures using confidentiality, integrity, availability, blockchain, and AI, gaps persist in securing the entire V2G ecosystem, particularly user behavior, energy market platforms, and vulnerabilities in advanced AI and physical access [57].

V2G boosts grid stability via EV power supply for load balancing, peak shaving, and frequency regulation. Yet, it introduces cybersecurity risks, with vulnerabilities to cyberattacks (Table 3 outlines the key threats associated with the V2G system) disrupting electricity flow. Robust security, including identifying weak nodes, network interdiction, optimized V2G fleets, and redundant pathways, is vital for resilience [58].

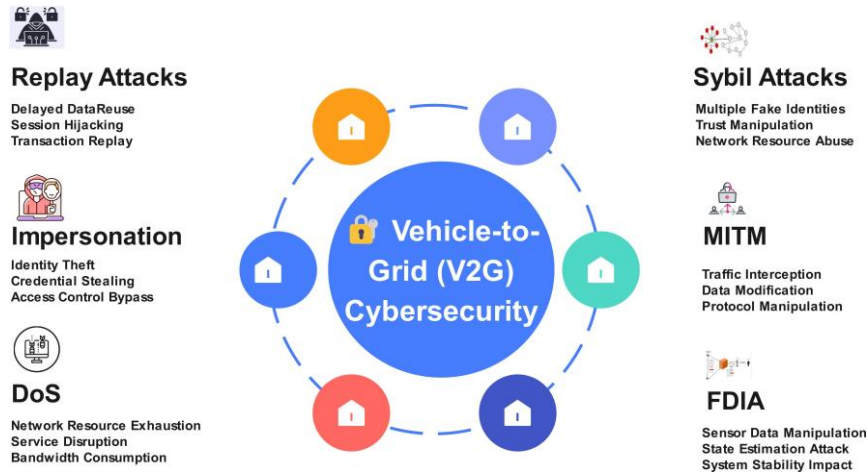

Figure 5: V2G Cyber Threat Classification

Electric vehicle network security presents multifaceted challenges across various communication domains [59]. V2G systems face complex cyberattacks including unauthorized access, data theft, and persistent threats—exploiting vulnerabilities in vehicle communications. To mitigate these risks, hybrid intrusion detection systems combining signature and anomaly-based methods with machine learning are crucial. Additionally, blockchain-NDN solutions enhance security by ensuring immutable data integrity and efficient content retrieval, demonstrating superior speed and latency in simulations. Future work focuses on real-world scalability testing [60, 61].

Emerging cybersecurity strategies leverage advanced technological approaches to mitigate these risks. Distributed edge computing emerges as a pivotal solution, enabling localized data processing and immediate erasure at charging points to minimize exposure risks [59]. Innovative authentication schemes, including pseudonym-based and conditional verification methods, further enhance privacy and security. Emerging technologies, such as distributed ledger systems and permissioned blockchain, offer additional layers of protection [62, 56].

### 3.1. Advanced Cybersecurity Solutions and Emerging Technologies

Various security solutions for Vehicle-to-Grid systems have evolved from traditional protocols to advanced AI-enhanced frameworks. Established security measures include the SUKA protocol, which addresses physical attacks such as man-in-the-middle, identity, and replay attacks through PUF for mutual authentication, enhanced by pseudonym IDs and dynamic cryptographic session keys [63]. Traditional data protection employs distributed edge computing for local data processing and federated learning for collaborative model training without raw data sharing [63].

The integration of Named Data Networking (NDN) and blockchain technology in V2G networks revolutionizes cybersecurity by replacing host-based addressing with efficient content-centric data retrieval. This approach addresses critical challenges like privacy risks and trust deficits through decentralized, tamper-proof ledgers. Key innovations include Proof-of-Authority (PoA) consensus for low-latency block validation and a Trie-based search algorithm that slashes content retrieval complexity from enabling rapid access to critical energy data [60, 61].

Core security infrastructure includes encryption, multi-factor authentication,

real-time monitoring systems, intrusion detection systems, and public key infrastructure (PKI) for secure message transmission [62, 56]. Building upon these foundational approaches, emerging cybersecurity frameworks increasingly integrate artificial intelligence and machine learning technologies to proactively address evolving cyber threats. Advanced V2G security frameworks implement sophisticated, pseudonym-free authentication mechanisms that ensure forward security, anonymity, and unlikability between electric vehicles and local aggregators (LAGs).

Current research emphasizes federated learning integration to preserve digital assets and user privacy, enabling charging stations to collaboratively train machine learning models without data sharing, thereby minimizing communication overhead and bolstering security [63]. The integration of distributed ledger technologies provides transformative solutions for enhancing transparency and trust through immutable, verifiable transaction records [62, 56].

Additional security implementations include permissioned blockchain models with covert channel authorization and group signatures for vehicle identity validation, complemented by behavioral analysis and cooperative vehicle detection systems [56]. Although, modern V2G security leverages a multi-faceted approach, combining established protocols like SUKA with AI-driven threat detection and federated learning for collaborative model training. Innovative authentication mechanisms and blockchain technologies further bolster security by ensuring anonymity, data integrity, and trust. The convergence of these advanced techniques provides a robust defense against evolving cyber threats within the V2G ecosystem.

Comprehensive security solutions for V2G integration have evolved from traditional protocols to advanced AI-enhanced frameworks. Established security measures include the SUKA protocol, which defends against physical attacks like man-in-the-middle, identity, and replay attacks using PUF for mutual authentication, bolstered by pseudonym IDs and dynamic cryptographic session keys [63]. Traditional data protection employs distributed edge computing for local data processing and federated learning for collaborative model training without sharing raw data [63].

Table 3: Key threats to V2G network

| Key Threats | Explanation | Solution | Ref |
|---|---|---|---|
| Address Resolution Protocol (ARP) Spoofing | Attackers pretend to be MAC address of trusted device and steal network traffic for eavesdropping on sensitive information or data tampering | Implement static ARP entries on critical devices and packet filtering for detection of suspicious traffic. Secure protocols like IPsec help by encrypting communications | [33] |
| Data Privacy | Sensitive user/energy data (e.g., usage patterns, pricing) exposed during transactions, risking exploitation. | Techniques such as ZKP that enable validation of transactions while keeping actual energy values and user identities confidential | [64] |
| Data Security Risks | V2G systems face cyberattack risks due to weak cybersecurity standards, potentially disrupting operations. Unauthorized data access could halt system functionality. | Implement TLS/SM4 encryption and E2MPS-S standards. Use secure, unified data platforms with real-time monitoring to counter threats | [65] |
| Impersonation Attack | Attack launched from garage charging station where bad actor hides charging station is masqueraded, putting user's safety and privacy at risk | Deploy mutual authentication measures where both EV and charging station claim the identity that is being presented | [17] |
| Side-Channel Attacks | Attacker can steal classified information from target using electromagnetic leakage or power consumption information | Use safe cryptographic techniques and insulated wires to reduce side channel data leakage | [66] |
| SQL Injection | Attackers introduce malicious SQL queries into databases for unauthorized access of data. | Adopt input tests and parameterized queries to prevent malicious SQL command execution. | [67] |
| Battery Manipulation Attack | EVs vulnerable to manipulations of charging parameters that can cause overcharge or over-discharge, rendering battery worthless or dangerous | Install diagnostic security modules (DSM) to monitor and regulate charging in real-time and stop abnormal usage automatically | [66] |
| Malware & Ransomware | Compromised EVSEs (charging stations) or BMS software can disrupt operations or demand ransom. | Regular firmware updates, sandboxing, and behavior-based threat detection to isolate and neutralize malware. | [68] |

Complementing these architectural advances, next-generation AI-driven security frameworks deploy enhanced Support Vector Machines and bio-inspired social spider optimization algorithms to detect CAN bus anomalies in real-time. For broader network protection, scalable IDS utilize fog-edge analytics to iden-tify novel threats in resource-constrained IoT environments typical of V2G ecosystems. Together, these layered solutions combining blockchain-enabled data integrity with adaptive machine learning defenses establish a robust security foundation for future smart grid infrastructures [60, 61].

Core security infrastructure includes encryption, multi-factor authentication, real-time monitoring systems, intrusion detection systems, and PKI for secure message transmission [62]. Current research stresses federated learning integration to protect digital assets and user privacy, allowing charging stations to collaboratively train machine learning models without sharing data, thus reducing communication overhead and strengthening security [63].

The use of distributed ledger technologies offers transformative solutions for boosting transparency and trust through immutable, verifiable transaction records [62, 56]. Additional security measures include permissioned blockchain models with covert channel authorization and group signatures for vehicle identity verification, along with behavioral analysis and cooperative vehicle detection systems [56, 57]. Although modern V2G security employs a layered approach combining established protocols like SUKA with AI-driven threat detection and federated learning for collaborative model training—innovative authentication methods and blockchain technologies further enhance security by ensuring anonymity, data integrity, and trust. The integration of these advanced techniques provides a resilient defense against evolving cyber threats within the V2G ecosystem.

### 3.2. Current Research on AI, ML, and DL Approaches to Secure V2G Networks

The contemporary landscape of cybersecurity for V2G networks increasingly depends on advanced technological (Fig.6 describes its role) methodologies in artificial intelligence, machine learning, and deep learning. Federated learning stands out as an innovative technique, facilitating collaborative development of machine learning models at charging stations while safeguarding data privacy and securing digital assets. By reducing communication overhead and enhancing data security, this approach ensures more efficient V2G operations. In partic-ular, the security challenges faced by driverless vehicles within 5G-enabled automotive networks are complex and multifaceted. These networks are vulnerable to significant threats such as eavesdropping, message fabrication, Sybil attacks, and other cyberattacks that could compromise communications between V2V, V2I, and V2G [63].

Researchers are enhancing vehicular cybersecurity with advanced intrusion detection systems, including improved SVM models for anomaly detection, AI-based deep learning engines for real-time attack classification, and hybrid systems achieving over 96% accuracy. Dynamic dataset frameworks integrate live threat intelligence to adapt to evolving attacks, while current models show high performance (95.6% accuracy), with future improvements targeting noise reduction and data quality [60]. AI and ML, including LSTM models and GA, optimize EV charging and routing by harmonizing grid demands with renewable energy utilization, thereby enhancing system efficiency. Deep Learning methodologies bolster security by identifying cyber threats through extensive data analysis. AI-driven models can reduce grid load by 27%, decrease load variation by 10%, and lower costs by 4%. Continued research is necessary to address challenges such as data privacy and the large-scale integration of EVS into power grids, such as that of Australia, with the objective of establishing secure and sustainable V2G systems [69].

## 4. Energy Market Dynamics and Business Models in V2G Integration

Energy market dynamics—such as consumer adoption rates, investment levels, and regulatory frameworks—play a decisive role in the large-scale deployment of V2G technology. The demand for V2G services is primarily shaped by fluctuations in electricity pricing, the availability of financial incentives for renewable energy integration, and competitive pressures among energy providers. With the accelerating adoption of V2G, the development of innovative business models and intelligent charging infrastructures has become essential to enhance scalability, operational efficiency, and economic viability. A detailed summary of major V2G projects, key collaborators, and implementation timelines is presented in Table 4.

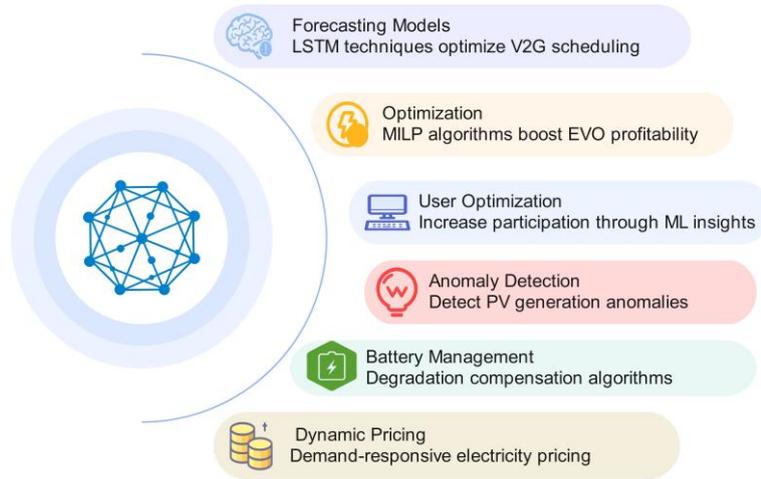

Figure 6: Role of AI and ML approaches in V2G system

The V2G framework offers significant financial benefits for prosumers, enabling them to generate revenue by selling excess stored energy to the grid or to local consumers. Financial incentives, including demand response programs and real-time pricing mechanisms, can yield considerable savings—estimated at up to $510 per month for participants. Moreover, V2G integration has been shown to reduce total energy costs by up to 7.8%, while savings on electricity purchases can reach 53.7% [70, 71, 72].

In the V2G ecosystem, electric vehicles act as mobile energy storage units, mitigating reliance on expensive peak-generation resources. They can absorb surplus renewable energy during periods of low demand and discharge it during high-demand intervals, thereby improving grid flexibility. P2P local electricity market models further enable efficient, decentralized energy trading, which reduces transmission losses and alleviates operational pressures on distribution system operators [73].

The V2G market leverages advanced trading mechanisms, including decentralized market-clearing systems that ensure privacy preservation and minimize power losses. Trading structures may involve bilateral agreements, day-ahead markets, and participation in ancillary services markets. The integration of emerging technologies such as AI and the IoT enhances the optimization of energy consumption patterns and operational forecasting [74, 75].

Table 4: Successful of implementation of V2G projects

| Project | Partner | Key Achievements | Location | Year | Ref |
|---|---|---|---|---|---|
| V2G power transition | State Grid Corporation of China's V2G project led by EV manufacturers | V2G cut power transition costs by 2.9 units and annual carbon emissions by 1.5 million tons, while optimizing peak load and reducing unused solar power | China | 2025 | [76] |
| Japan Renewable Grid Optimization Project | Energy Companies, like TEPCO, for grid integration | The V2G project in Japan optimized power output to 29.1% solar and 37% renewables for under $2 million, enhancing grid integration with a 22.7% to 37% renewable energy share and enabling 30 MWh of resalable energy. | Japan | 2025 | [77] |
| PG&E V2G pilot program | BMW, Ford, Honda, Nuvve | California integrated 100 EVs, providing 1.2 MW of stored energy during peak demand, enabling participants to earn $800 annually through energy trading. | United States | 2025 | [78] |
| INEES Project | Volkswagen and Fraunhofer ISE | Demonstrated that a fleet of 50 EVs could deliver 2 MWh of grid storage capacity, reducing electricity costs by 20% for consumers. | Germany | 2025 | |

Globally, more than 134 V2G projects have been launched, with Europe and North America leading in adoption, reflecting growing market confidence and maturity [79]. Nevertheless, successful deployment depends on several key factors: user acceptance, battery availability, EV adoption rates, and charger power capacity. Barriers remain, including concerns over battery degradation, uncertainties regarding long-term economic viability, and data privacy risks [80]. For V2G technology to achieve mainstream penetration, strategies must focus on consumer engagement, incentive structures, and innovative market-driven mod-els. The combination of tangible financial benefits—such as revenue from energy sales and reduced dependency on peak power plants—and advancements in trading and control technologies positions V2G as a transformative element in modern energy systems. Overcoming existing challenges will be critical to unlocking its full potential within global energy ecosystems.

*4.1. Market Challenges and Potential of V2G in Reducing Peak Demand*

The widespread deployment of V2G technology faces several market-related challenges that directly influence its effectiveness in reducing peak electricity demand. One of the most critical factors is user participation. The willingness of EV owners to engage in V2G programs is essential; however, participation rates may be limited by a lack of awareness or insufficient understanding of the financial and environmental benefits. Battery degradation is another major concern. Frequent charging and discharging cycles required for energy trading can shorten battery lifespan and degrade performance, thereby raising questions about the long-term economic viability of participation. Additionally, the absence of supportive regulatory frameworks can hinder adoption. Without clear policies, incentives, and standardized procedures, both consumers and utilities may be reluctant to commit to large-scale V2G integration.

Market dynamics are further shaped by electricity price fluctuations, which can influence consumer interest. Attractive electricity pricing models—such as real-time pricing or peak-period compensation—are necessary to encourage participation. Despite these challenges, V2G technology offers substantial potential for peak demand reduction. By enabling EVs to discharge stored energy back to the grid during high-demand periods, V2G can enhance grid stability, improve system efficiency, and provide direct financial incentives to EV owners. These benefits align with the broader transition toward renewable energy integration and decentralized energy systems. Globally, there are currently 134 V2G projects, with the majority located in Europe and North America [79]. The market is expanding rapidly—valued at $5,259.87 million, it is projected to reach

$7,373.29 million by 2025 and a staggering $109,938.44 million by 2033, representing a compound annual growth rate (CAGR) of 40.18%. The rapid adoption of EVs and smart grid technologies is the primary driver of this growth, as illustrated in Fig.7, which also highlights regional variations. Moreover, Fig.8 presents market forecasts for both unidirectional and bidirectional V2G, revealing different growth trajectories over time. The future success of V2G will depend on targeted strategies to:

1. Overcome user hesitancy through awareness campaigns and consumer education.

2. Mitigate battery degradation via technological improvements and optimized charging algorithms to ensure economic sustainability.

3. Establish supportive regulatory frameworks combined with dynamic electricity pricing schemes to incentivize participation.

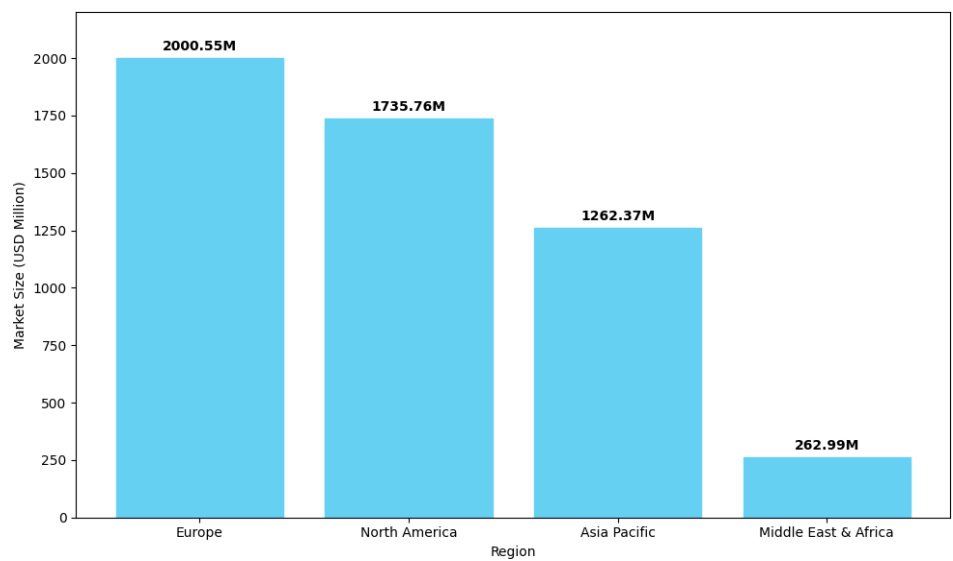

Figure 7: V2G Market Forecast (2025–2033) [81]

If these barriers are addressed, the combination of rising renewable energy adoption, supportive policy environments, and increasing numbers of global projects positions V2G as a transformative tool for peak demand management and long-term grid resilience.

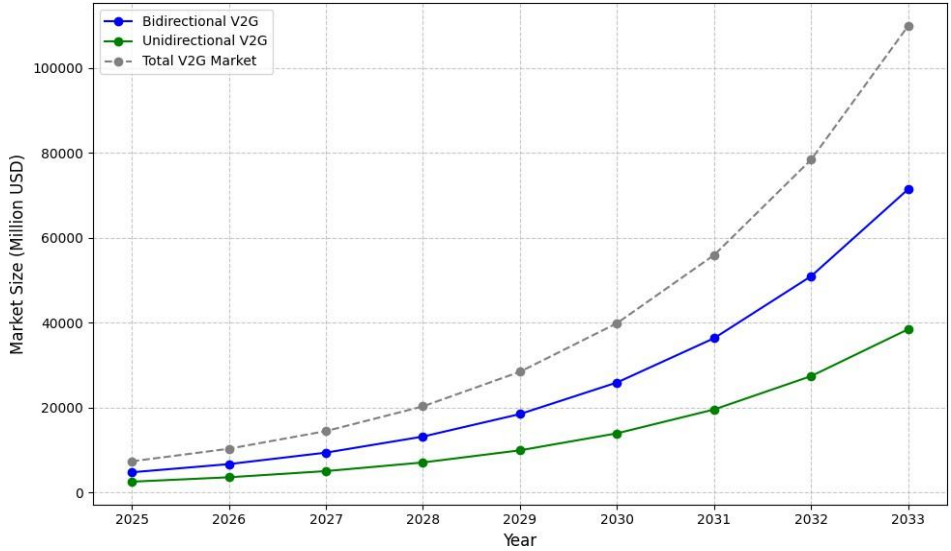

Figure 8: V2G Market Forecast (2025–2033) [81]

## 4.2. Regulatory and Policy Frameworks for V2G Implementation and Grid Stability

The effective implementation of V2G technology is highly dependent on the regulatory and policy environment that governs its adoption and integration, as illustrated in Fig. 9. Four primary regulatory components—value frameworks, grid codes, connection agreements, tariffs, and market platforms—have been identified as pivotal in facilitating the use of flexibility at the distribution level [12]. These frameworks, tested in various demonstration projects with EV fleets, aim to ensure the seamless integration of V2G into existing power systems.

Regulatory approaches vary significantly across regions, influencing both consumer participation and investment strategies. Favorable government incentives, particularly those promoting renewable energy integration, enhance the economic viability of V2G by enabling EV owners to sell surplus energy back to the grid or local consumers. The growing demand for V2G services is also fueled by competition among energy providers and rising consumer expectations for sustainable energy practices.

Blockchain technology presents an opportunity to strengthen these frameworks by enabling transparent transactions and lowering operational costs. However, challenges remain, including the need for supportive regulations that address concerns about battery degradation, cost-effectiveness, and data privacy.

A notable issue is High Demand Disparity, where V2G systems may be under-utilized or overloaded—especially in regions with high EV penetration, such as Tokyo and Aichi, Japan. The successful deployment of V2G technology thus relies on a strong regulatory landscape that encourages innovation while addressing barriers to adoption [82, 83, 84].

International examples highlight diverse regulatory strategies. In China, the evolving V2G policy framework demonstrates a strong commitment to sustainable energy transformation. The State Grid Corporation's USD 350 billion investment supports comprehensive V2G deployment, renewable integration, and grid modernization, aimed at reducing coal dependence while improving efficiency. In South Korea, government policies promote EV infrastructure and V2G adoption, reflected in the sale of nearly 10,000 FCEVs in 2020. The national target of 600,000 EV charging plugs by 2030 underscores a strategic approach to enhancing grid stability and renewable energy integration [85, 77].

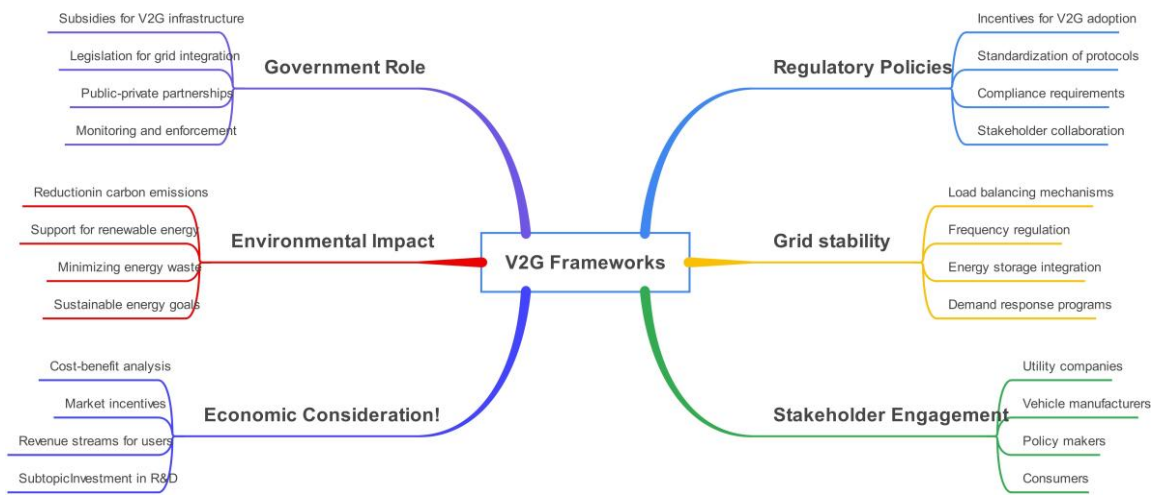

Figure 9: Conceptual model of V2G frameworks and Polices

To realize V2G's full potential, comprehensive regulatory reforms are essential. Key measures include [86, 87]:

1. Creating new ancillary service markets and fair compensation mechanisms for V2G services.

2. Updating grid codes to enable bidirectional energy flow and leverage V2G for grid stability.

3. Removing barriers to energy storage participation.

4. Streamlining energy transaction guidelines to allow EV owners to sell surplus energy without excessive administrative hurdles.

Regulatory bodies must also offer consumer and utility incentives to encourage participation. Prosumer savings can reach up to €510 per month [70], and blockchain-enabled decentralized trading mechanisms can enhance privacy, reduce power losses, and improve market efficiency [71, 74].

As adoption accelerates, addressing these regulatory challenges will be critical to maximizing V2G's role in a flexible, resilient, and sustainable energy market. Well-designed frameworks will not only accommodate the bidirectional energy exchange between EVs and the grid but also adapt to evolving consumption and production patterns, driving a successful transition toward low-carbon energy systems [83].

*4.3. Business Models and Market Participation*

V2G technology supports a range of business models, with P2P energy trading and utility–grid partnerships being the most prominent. In the P2P model, EV owners (prosumers) sell surplus stored energy directly to local consumers or the grid, fostering vibrant local energy markets and optimizing renewable energy utilization. This approach empowers prosumers and offers substantial financial rewards, with potential savings reaching €510 per month[4].

Utility–grid partnerships leverage existing grid infrastructure to enhance stability by balancing supply and demand during peak hours. In these arrangements, EVs operate as mobile energy storage units, enabling flexible energy dispatch. Financially, V2G can reduce electricity prices by up to 7.8% and cut grid power purchase costs by up to 53.7% [72], delivering tangible benefits for energy industry stakeholders, policymakers, and researchers.

Decentralized trading mechanisms further strengthen these models by improving user privacy, minimizing power losses [74], and reducing operational strain on distribution system operators. By integrating such mechanisms, V2G business models not only support grid stability but also contribute to the broader goal of sustainable energy transitions. The market potential of V2G stems from its dual role in energy markets and backup power services, enabling profitable energy sharing among users while supporting grid reliability [84]. Achieving this potential requires addressing regulatory barriers and fostering innovative market structures.

A comprehensive market analysis—presented in Table 5—outlines the technical, financial, and operational factors critical to V2G success. These include cost control strategies, advanced charging infrastructure, renewable energy integration, and novel business models. The deployment of modern technologies such as real-time energy management and decentralized trading systems enhances scalability, dependability, and overall system sustainability. By combining innovative business models with supportive regulatory frameworks and advanced technologies, V2G can fulfill its promise as a grid-stabilizing, revenue-generating, and sustainability-driving solution for the evolving energy market.

Table 5: Successful of implementation of V2G projects

| Location | Project | Market Analysis | Year | Ref |
|---|---|---|---|---|
| China | Chinese State Grid Corporation and general SG projects | Chinese State Grid Corp. and SG projects target 30% renewable energy by 2025 with $350B in Smart Grid investments, driven by policies tackling high costs, technical issues, and reducing energy losses and operational costs | 2025 | [85] |
| Japan | V2G pricing analysis | V2G market grows with rising demand and 37% renewable target by 2030, 1.36M EVs, and 57.47 yen/kWh pricing, offering 30,000 MWh resale. High charger costs and standards gaps hinder adoption, but V2G boosts grid stability | 2024 | [77] |
| Germany | PowerACE | Agent–based load scheduling to optimize V2G EV dispatch in Germany, showing tariff structures and battery aging affect profitability, while reducing stationary storage needs by 8 GW (2034 projection) | 2024 | [88] |

Table 5: Successful of implementation of V2G projects

| Location | Project | Market Analysis | Year | Ref |
|----------|---------|-----------------|------|-----|
| Europe (based on 7 European Countries) | Report on consumer behavior | Report: Self–sufficiency, environment, and incentives boost V2G adoption. Most EV users join if costs, renewables, and stability are secured. Barriers include battery wear, load management, and low trip loads | 2023 | [89] |
| South Korea | V2G trends and po-lices | V2G benefits from 10,000 FCEVs and 9% hydrogen stations glob-ally. 2.9% EV sales with incentives support integration despite battery degradation. EV2030 targets major renewable energy by 2030 | 2022 | [90] |
| Europe (based on 7 European Countries) | Stakeholder analysis | Self-sufficiency, environment, and incentives boost V2G adoption. Most EV users join if costs, renewables, and stability are ensured. Barriers include battery wear, load management, and low trip loads | 2022 | [91] |
| US | General Motors (GM) | GM's $35B investment solidifies its role in the global EV market, prioritizing battery tech, charging networks, and V2G. Despite supply chain and adoption hurdles, its Ultium platform and rapid model expansion target market leadership by 2025 | 2021 | [92] |

## 4.4. Market-Driven Incentives Supporting V2G Adoption

The adoption of V2G technology is largely influenced by market-driven incentives designed to stimulate both investment and consumer participation. One of the most compelling benefits for EV owners is the opportunity to generate up to €510 per month by selling surplus energy back to the grid [70], providing a strong financial motivation for participation. Blockchain integration enhances these market mechanisms by enabling transparent transactions, reducing operational costs, and increasing trust among stakeholders [71]. The rise of peer-to-peer (P2P) local electricity markets represents a transformative shift in energy trading [73, 74]. Through these decentralized networks, EVs can strategically absorb excess renewable energy, effectively balancing supply and demand while minimizing transmission losses. Privacy-preserving trading mechanisms embedded in these models further reduce the operational burden on traditional distribution networks.

As mobile energy storage units, V2G-enabled EVs deliver tangible benefits to the grid, particularly in reducing overall electricity costs by up to 7.8% and improving grid stability through strategic energy provision during peak demand periods [80]. However, achieving widespread adoption requires addressing several critical barriers:

1. Battery degradation concerns, which affect long-term economic viability.

2. Robust data privacy protections to maintain user trust.

3. Economic feasibility assessments to ensure sustained participation.

Regional differences also influence market effectiveness. For example, in Japan, tailored strategies are required to address High Demand Disparity, where V2G utilization levels can vary significantly across regions such as Tokyo and Aichi[82]. Innovative pricing strategies—such as auction-based methods and blockchain-enabled settlements—can further enhance transparency, optimize market operations, and build confidence among participants [83]. Overall, the market outlook for V2G is highly promising, as it enables EVs to support grid stability, strengthen local energy networks, and contribute directly to the achievement of low-carbon energy goals[84].

## 5. Discussion

### 5.1. Interconnections between Grid Stability, Cybersecurity, and Market Dynamics

V2G integration adoption is primarily driven by market forces, with innovative incentive schemes designed to encourage investment and consumer engagement. By selling excess energy back to the grid, consumers may earn up to €510 per month, representing a substantial financial benefit. By establishing transparent transaction methods and significantly lowering operating costs, the integration of blockchain technology further maximizes market efficiency. The emergence of P2P local electricity market models represents a transformative approach to energy trading. EVs can strategically absorb excess renewable energy through these decentralized networks, which effectively balance supply and demand. By minimizing transmission losses and promoting privacy-preserving trading mechanisms, such models significantly alleviate the operational burdens on traditional distribution networks. Fig.10 demonstrates V2G systems integrated framework and its remarkable potential as mobile energy storage solutions, offering tangible benefits to the electrical grid.

Research indicates these systems can potentially reduce overall electricity costs by up to 7.8%, providing hope for significant economic benefits. This reduction of expenses is coupled with the enhancement of grid stability through strategic energy provision during peak demand periods. Nevertheless, widespread adoption remains contingent upon addressing critical challenges, including concerns about battery degradation, economic feasibility assessments, and robust data privacy protections. In Japan, tailored strategies are necessary to manage the "High Demand Disparity," with regional factors influencing the effectiveness of V2G in achieving low-carbon goals. Optimizing pricing strategies through auction methods and blockchain can enhance transparency and trust in V2G systems. Overall, the market outlook for V2G is promising, as it allows EVs to stabilize the grid and support local energy setups.

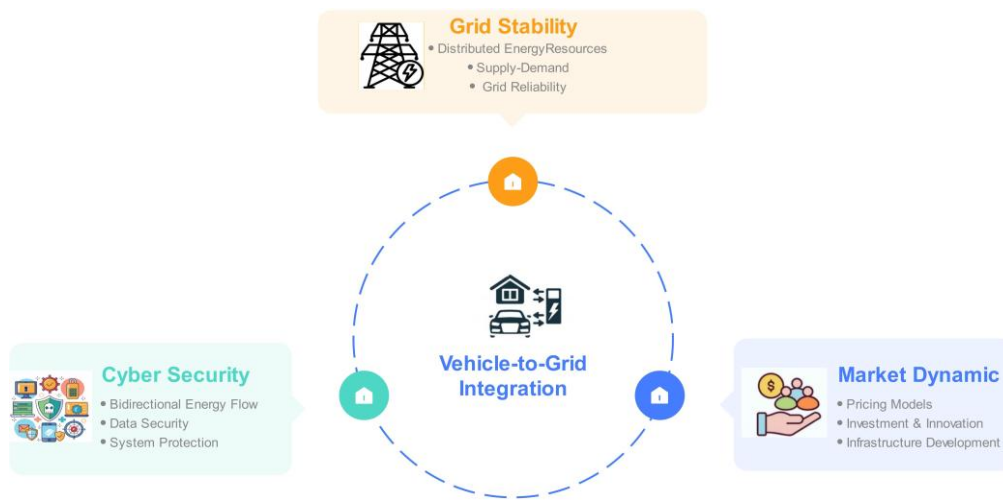

Figure 10: Integrated Framework for Vehicle-to-Grid (V2G) Integration

1. **_Grid Stability:_** V2G technology offers several avenues for enhancing grid stability. By responding swiftly to frequency fluctuations, EVs can inject or absorb power, maintaining the grid's frequency within acceptable lim-its. Furthermore, strategically positioned V2G systems can provide reactive power, bolstering voltage levels, particularly in areas with high renewable energy penetration or extensive distribution networks. V2G can also contribute to peak shaving by discharging EV batteries during periods of high electricity consumption, mitigating the need for costly peak-generating power plants and improving overall grid reliability. EVs equipped with V2G capabilities could potentially offer black start services, aiding in the restoration of power to critical infrastructure following a blackout. However, managing a large fleet of EVs for grid services poses a significant technical challenge. This challenge underscores the need for sophisticated control systems and communication infrastructure to ensure the seamless aggregation and dispatch of energy from numerous vehicles, highlighting the complexity of the task.

2. ***Cybersecurity:*** The utmost importance as V2G systems become increasingly integrated into the grid. Securing individual EVs and charging stations requires robust authentication, authorization, and access control mechanisms to prevent unauthorized access and control. Protecting communication channels between EVs, charging stations, grid operators, and aggregators is also essential, employing encryption, secure protocols, and intrusion detection systems to prevent eavesdropping and tampering. V2G systems generate substantial data regarding energy consumption, driving patterns, and user behavior, making data security and privacy paramount to maintain user trust and comply with privacy regulations. Moreover, ensuring the security of hardware and software components used in V2G systems is crucial, as attackers can exploit vulnerabilities to compromise the entire system. Given the constantly evolving cybersecurity threat landscape, continuous monitoring, assessment, and adaptation of security measures are imperative.

3. **Market Dynamics:** Market dynamics are pivotal in shaping the adoption and economic viability of V2G technology. Participants can leverage value stacking to generate revenue from multiple sources, such as providing frequency regulation, peak shaving, and energy arbitrage, optimizing the value stack to maximize profitability and attract investment. Clear and appealing incentive structures are necessary to encourage EV owners to participate in V2G programs, potentially including direct payments, reduced electricity rates, or other financial benefits. The design of electricity markets must accommodate the unique characteristics of V2G, such as the intermittent availability of EVs and the need for real-time dispatch. Allowing competition among V2G aggregators can foster innovation and improve service quality, but it is equally important to ensure that aggregators have sufficient scale and resources to effectively participate in the market. The V2G market is still nascent and faces considerable uncertainty regarding technology costs, regulatory policies, and consumer adoption, making the management of this uncertainty and mitigation of risks essential to attract investment and ensure long-term sustainability.

*5.2. Cybersecurity Vulnerabilities Impacting Grid Reliability and Market Models*

The integration of vehicle-to-everything (V2X) technologies and telematics systems into fleet management and grid services introduces substantial cybersecurity vulnerabilities that threaten both grid reliability and emerging market models. These technologies, including V2G programs that generate approximately $2,000 per vehicle annually and telematics systems that cost around $500 per vehicle for installation, depend on interconnected devices and software, thereby creating potential entry points for cyberattacks. Such vulnerabilities could disrupt grid operations, such as load balancing in pilot programs in Georgia and Virginia, and undermine confidence in market frameworks, thereby impacting the projected $65.84 billion fleet management market by 2023. As EVs

adoption increases, with energy cost savings of 40-60% in regions like California, unsecured systems pose risks that could destabilize grid reliability and impede market expansion. It's crucial to maintain market growth by addressing these vulnerabilities through comprehensive cybersecurity measures.

1. ***Cybersecurity Vulnerabilities:*** The integration of V2X and telematics systems into fleet operations introduces significant cybersecurity vulnerabilities due to their dependence on connected devices and software. However, these systems also offer the potential for substantial energy cost savings. Systems like GPS trackers, OBD-II devices, and advanced telematics, with installation costs ranging from $100 to $500 per vehicle, are vulnerable to hacking risks, including data interception and unauthorized access to vehicle controls. For instance, complex telematics systems requiring professional installation, often costing $50–200 per vehicle, may involve hardwired setups that, if not properly secured, could expose critical fleet and grid data. The fleet management market is expected to grow to \$65.84 billion by 2023, driven by the increased adoption of these technologies, which expands the attack surface for cyber threats. In regions with high EV penetration, such as California, where pilot programs have demonstrated energy cost savings of 40-60%, unsecured systems could compromise sensitive grid interaction data. Robust cybersecurity measures, including end-to-end encryption, secure APIs, and regular software updates, are essential to safeguard these systems and prevent potential breaches that could disrupt fleet operations and grid services.

2. ***Impact on Grid Reliability:*** Cybersecurity risks in V2X systems, especially V2G programs, threaten grid stability by potentially disrupting energy flow management. The electric school buses in Georgia and Virginia generate $1,000–2,000 annually through V2G programs, depending on real-time data for services like peak shaving and load balancing. However, a cyberattack

could manipulate energy distribution, risking grid instability or outages, especially in high EV adoption areas such as California, where per-mile energy costs have decreased by 40-60%. This could lead to significant economic loss. The dependence on connected devices for instant communication between vehicles and grid operators heightens this risk; a breach could disrupt vital data exchanges or enable malicious commands. For instance, compromised telematics could affect the coordination necessary for grid services, jeopardizing the stability demonstrated in pilot programs from 2020 to 2023. To protect grid reliability and ensure continuous energy service, implementing secure communication protocols and intrusion detection systems is essential.

3. ***Impact on Market Models:*** Cybersecurity vulnerabilities in V2X and telematics systems can significantly disrupt market models by eroding trust in revenue-generating opportunities, such as those from V2G programs and telematics subscriptions. V2G programs in Virginia and Georgia generate $1,000–1,600$ per vehicle annually, while telematics subscriptions cost $15–40$ monthly, supporting the fleet management market's projected growth at a 20.07% CAGR. However, a cyber breach could deter fleet operators from participating in grid services, as they fear data breaches or financial losses, thereby reducing these revenue streams. Additionally, the 7% rise in U.S. fleet maintenance costs in 2024, driven by parts shortages and the adoption of EV technology, could be exacerbated by expenses related to cyber-attack recovery, particularly for smaller operators. This could slow market growth, as operators may hesitate to invest in vulnerable technologies. Ensuring robust cybersecurity through secure software integration and regular system audits is vital to sustaining confidence in these emerging markets and supporting their long-term viability. This emphasis on cybersecurity should reassure stakeholders about the future of the market.

### 6. Challenges in implementing V2G integration

Managing power fluctuations in distribution systems becomes increasingly complex with the integration of V2G. This innovative approach introduces new challenges in maintaining grid stability, ensuring cybersecurity, and controlling infrastructure costs. Addressing these concerns requires a comprehensive strategy that incorporates system optimization techniques, predictive analytics, and robust cybersecurity frameworks to protect vulnerable points. The following section examines several key challenges posed by V2G implementation:

#### 6.1. Grid Stability and Load Management

1. Grid Operations and Voltage Control: Sustaining operational effectiveness across fluctuating grid conditions remains a critical challenge in V2G implementation. Managing voltage equilibrium in NPC (Neutral Point Clamped) architectures requires sophisticated monitoring and control systems to ensure stable power delivery despite variable grid conditions [93]

2. Infrastructure and Real-Time Coordination: The financial implications and technical intricacy of durable matrix converters present significant barriers to widespread adoption [93]. Orchestrating decentralized energy resources in real time demands sophisticated control mechanisms and robust communication frameworks to facilitate seamless V2G implementation across distributed networks.

3. System Scalability and Economic Viability: Developing expandable, economically viable approaches capable of accommodating grid inconsistencies while navigating system complexity represents a fundamental challenge[94]. These technical hurdles emphasize the necessity for scalable solutions that can maintain operational integrity across varying grid conditions while ensuring long-term economic sustainability.

#### 6.2. Scheduling methods to enhance Battery life

1. Computational Complexity and System Responsiveness: The coordination

of power fluctuations within V2G networks presents multiple obstacles, particularly the intensive processing requirements of advanced optimization techniques such as PSO, DE, WOA, and GWO. These computational demands can compromise system responsiveness as EVs deployments grow, creating scalability concerns for large-scale implementations.

2. Real-Time Market and Load Response: Effectively responding to instantaneous pricing signals and shifting load patterns poses significant difficulties for V2G systems. The dynamic nature of electricity markets necessitates continuous system adjustments to preserve grid reliability and economic efficiency, requiring sophisticated algorithms capable of rapid decision-making under changing operational conditions.

3. Battery Degradation and Economic Impact: The repeated charging-discharging cycles inherent in V2G operations contribute to accelerated battery degradation, resulting in elevated long-term expenses for EVs owners [95]. This degradation not only affects the economic viability of V2G participation but also raises concerns about the overall lifecycle costs and sustainability of EVs investments in grid services.

*6.3. Scalability and Latency*

1. System Performance: Coordinating multiple concurrent transactions while ensuring minimal delays and maximum processing capacity continues to present a formidable challenge in V2G systems. The simultaneous management of numerous vehicle-to-grid interactions requires robust computational infrastructure capable of handling high-volume, real-time data processing without compromising system performance.

2. Security Protocols: Current protocols, including Emularis, need substantial reinforcement to withstand newly developing cybersecurity threats and guarantee ongoing system resilience[96]. The evolving threat landscape necessitates continuous updates and strengthening of security measures to protect critical V2G infrastructure from potential vulnerabilities and malicious attacks.

3. Infrastructure Innovation: These obstacles highlight the critical requirement for sophisticated technological innovations capable of handling the increasingly intricate nature and expanding requirements of V2G network infrastructures [96, 97]. The growing complexity of modern grid systems demands advanced solutions that can scale effectively while maintaining operational integrity and security across diverse deployment scenarios.

## 6.4. Privacy Preserving in V2G Integration

1. Trustworthiness and System Dependability: Vehicle-to-Grid systems face challenges in establishing trust due to their decentralized and dynamic data communication patterns. Ensuring safety, security, and operational predictability is critical for reliable performance. The lack of standardized protocols for secure and consistent energy transfer complicates integration into critical infrastructure networks [98].

2. Privacy Vulnerabilities:The frequent exchange of sensitive data across multiple vehicle connection points in V2G systems creates significant privacy risks. These interactions increase exposure to potential security breaches, requiring robust safeguards to protect user information and maintain system integrity [99].

3. Data Integrity and Security Threats:V2G systems are susceptible to data poisoning and other attacks that threaten data integrity. Implementing strong protective measures is essential to ensure dependable system operation and prevent compromises that could disrupt energy transfer processes [100].

## 6.5. Secure Communication

1. Vulnerability to Attacks: V2G systems face significant security risks, including impersonation tactics, replay attacks, and man-in-the-middle interceptions. These threats exploit the dynamic nature of V2G communication, requiring advanced and adaptable defensive strategies. Current security protocols often fall short in addressing these evolving risks, leaving systems exposed to potential breaches [96].

2. Complexity of Dynamic Wireless Charging:The introduction of dynamic wireless charging technology adds further complexity to V2G security. This technology demands specialized protection mechanisms to secure simultaneous bidirectional data and energy exchanges. Sophisticated, tailored cryptographic approaches are essential to mitigate these vulnerabilities and ensure the dependability and trustworthiness of V2G infrastructures [96].

## 6.6. Market Viability and Interoperability Analysis

1. Revenue Model Challenges: Developing sustainable revenue frameworks for V2G, such as subscription-based or usage-dependent payment systems, is hindered by unpredictable grid demand fluctuations. These inconsistencies make it difficult to establish stable and profitable business models for V2G integration [101].

2. Interoperability and Standardization Issues: Inconsistent protocols across regions and manufacturers create significant barriers to seamless V2G system integration. Variations in regulatory approaches and communication standards, as shown in Table 6, further complicate efforts to achieve unified standards for interoperability [101].

3. Evolving Security Threats:The evolving landscape of security vulnerabilities, including threats like eavesdropping, demands sophisticated protective strategies. Robust measures are essential to safeguard system integrity and ensure the reliability of V2G infrastructures [102].

Table 6: Regulatory concerns in several regions

| Region | Regulatory Challenges | Standardization Issues | Ref |
|--------|----------------------|------------------------|-----|
| Canada | The need for the establishment of state and federal EV infrastructure leads to a variety of regulated states. There are some areas where there is no incentive for private investment. | There are inconsistent plug and charge standards, leading to problems with interoperability among charging stations. | |
| USA | Funding and implementation efficiency are impacted by fragmented laws resulting from complex federal and state policies. Laws pertaining to EV security and data privacy are currently being developed. | Various standards (CCS, Chademo, etc.) are used on interstates, leading to user and manufacturer compatibility issues. | [101] |
| South Africa | Limited government incentives for EV adoption, high electricity prices affecting charging affordability, and a slow rollout of public charging infrastructure. | Dual adoption of CHAdeMO and CCS charging standards, absence of interoperability between charging networks, and the necessity of standardized grid integration policies. | |
| China | Low participation in centralized V2G operations highlights the need for regulatory incentives or mandates. Effective V2G integration and scalability also require stakeholder coordination and clear regulatory frameworks to define roles and data-sharing protocols. The absence of such policies could be a hindrance. | Standardization ensures V2G compatibility across EVs, charging infrastructure, and grid systems. Uniform protocols for cellular/Wi-Fi communication and user interfaces enable seamless operation and scalability across diverse EV models. | [103] |

## 6.7. Regulatory Framework and Policy Formulation

1. Cybersecurity and Data Privacy: Vehicle-to-Grid integration encounters significant regulatory hurdles in ensuring cybersecurity and data privacy. The integration of 5G with V2G systems necessitates robust encryption protocols to protect sensitive data and maintain system security, addressing the complex needs of various stakeholders [101].

2. Energy Bidirectionality Regulations: The bidirectional nature of energy flow in V2G systems introduces regulatory challenges that require clear guidelines. Establishing explicit rules for energy exchange is essential to ensure compliance and operational efficiency across diverse regulatory environments [101].

3. Collaboration and Legal Frameworks: Effective coordination among stakeholders in V2G and 5G integration demands clear legal frameworks to foster trust and security. These frameworks must address inter-operator collaboration and data exchange challenges to support seamless and secure V2G system operation [102].

## 7. Recommendations for Future Research

Vehicle-to-Grid facilitates bidirectional energy exchange between EVs and the electrical grid, providing advantages including grid stabilization, enhanced renewable energy integration, and financial benefits for EV owners. Despite these promising capabilities, numerous technical and operational challenges require resolution to achieve widespread V2G deployment. Below are key future research areas:

### 7.1. *Ensuring Grid Stability Through Effective Load Management*

1. Enhancing Grid Stability: The use of wide band-gap materials like SiC and GaN in V2G systems enhances scalability and efficiency. These materials outperform traditional technologies, supporting a more reliable and effective V2G framework.

2. Real-Time Responsiveness with Edge Computing: Distributed edge computing enables real-time responsiveness to power demand fluctuations in V2G systems. This technology improves load management by facilitating rapid and precise adjustments to grid requirements.

3. AI-Powered Resource: AI-powered control systems optimize the coordination of distributed resources in V2G frameworks. These systems reduce operational overhead and enhance reaction accuracy, contributing to a more scalable and stable grid.

### 7.2. *Optimized Scheduling Strategies for Prolonging Battery Life*

1. Hybrid Algorithmic Frameworks: Hybrid algorithmic frameworks enhance scalability and efficiency in V2G network infrastructures. These frameworks optimize system performance by integrating advanced computational techniques for robust operation.

2. AI-Powered Real-Time Scheduling: AI-driven solutions enable dynamic scheduling adjustments in response to grid demand and market conditions. This real-time optimization improves V2G system responsiveness and operational efficiency.

3. Advanced Battery Management: Sophisticated battery management protocols minimize degradation and extend battery longevity in V2G systems. These protocols reduce financial burdens for users, enhancing cost-efficiency and system durability.

*7.3. Optimizing Scalability and Latency*

1. Quantum-Resistant Cryptographic Protocols: Implementing quantum-resistant cryptographic protocols strengthens V2G system resilience against advanced cyber threats. These protocols enhance security, ensuring robust protection for critical network operations.

2. Decentralized Transaction Management: Decentralized architectures, integrating blockchain with Emularis frameworks, provide scalable and secure transaction management for V2G systems. These solutions enhance efficiency and adaptability in handling transactions.

3. Adaptive Distributed Payment Infrastructures: Distributed payment infrastructures with adaptive capabilities address V2G's scaling and security challenges. These systems establish a robust foundation, preparing networks for future operational demands and growth.

*7.4. Enhancing Trustworthy and Secure V2G Systems*

1. Edge-Based Frameworks for Reliability: Prioritizing edge-based frameworks in V2G systems enhances reliability through secure, real-time operations. These frameworks ensure dependable performance, supporting seamless integration with critical infrastructure.

2. Advanced Privacy-Preserving Technologies: Implementing homomorphic encryption, differential privacy, and federated learning safeguards sensi-tive data across V2G vehicle endpoints. These technologies protect privacy while maintaining secure data exchanges.

3. Data Integrity with Federated Learning: Using federated learning alongside grouping techniques like HE-NFRT, Mask-NFRT, or Krum mechanisms prevents data poisoning in edge-level processing. These methods ensure data integrity, optimizing V2G system security and efficiency.

### 7.5. Strategies for Robust V2G Communication

1. Advanced Cryptographic Techniques: Modern V2G systems must adopt advanced cryptographic methods, such as ECC combined with quantum-resistant algorithms, to ensure long-term security. These techniques provide robust defenses against evolving cyber threats.

2. Flexible Security for Wireless Charging: Wireless charging in V2G systems requires adaptable security solutions to protect bidirectional data and power exchanges. Tailored protocols are essential to address the unique challenges of these dynamic interactions.

3. Scalable and Forward-Compatible Infrastructure: As V2G networks grow in complexity and demand, developing expandable and forward-compatible infrastructure is critical. Evolving security protocols alongside emerging technologies ensures resilience against sophisticated threats.

### 7.6. Interoperability and Market Viability

1. Standardized V2G Regulations: Collaboration between governments and the private sector is essential to establish standardized V2G rules for seamless EVs integration across diverse grids. These standards promote sustainability, reduce carbon emissions, and create revenue opportunities for the energy sector.

2. Interoperability: Projects like 5GCAR, 5G-DRIVE, 5GCroCo, and VITAL-5G are critical for achieving interoperability and cross-border alignment in V2G systems. These initiatives drive scalable solutions and validate investments, fostering robust business models for global integration.

3. Security and Scalability with Machine Learning: Combining machine learning with physical layer security addresses emerging threats in V2G networks. These advancements enhance system security, scalability, and worldwide integration, ensuring resilient and efficient V2G operations.

*7.7. Strategies Related to Regulation and Policy*

1. Regulatory Frameworks: Developing regulations that enable bidirectional energy exchange while aligning with cybersecurity and data privacy policies is crucial. These frameworks build confidence in V2G systems, ensuring secure and reliable adoption.

2. 6G for Reliable Communication: The deployment of 6G networks will provide ultra-low latency and dependable communication for real-time V2G operations. This advancement enhances system responsiveness and supports efficient energy transfer.

3. Enhancing Efficiency with Machine Learning and Digital Twins:Integrating machine learning with digital twins improves the operational efficiency and resilience of V2G ecosystems. These technologies optimize system performance, paving the way for scalable and secure V2G adoption.

## 8. Conclusion

This review highlights that the successful deployment of V2G integration rests on the integration of three critical pillars: grid resilience, cybersecurity, and market design. Technically, V2G contributes to grid stability by supporting frequency regulation, voltage control, and load balancing. However, integration challenges persist, including infrastructure scalability, battery degradation, and harmonic distortion. Emerging solutions such as AI-driven forecasting, BESS, and adaptive scheduling algorithms are essential for mitigating these technical barriers. Cybersecurity remains a significant concern, with threats such as impersonation, data manipulation, and system intrusion requiring robust, multi-layered defense mechanisms. Techniques such as PUF-based authentication, fed-erated learning, and blockchain-enabled trust frameworks offer promising protection for V2G networks. On the economic front, P2P energy trading and dynamic pricing models present new revenue streams and user engagement opportunities. However, widespread adoption is hindered by fragmented policies, a lack of standardized protocols, and insufficient consumer incentives. To fully realize the potential of V2G systems, it is imperative to scale real-world pilot projects, promote protocol harmonization, and strengthen cross-sector collaboration. Aligning technological capabilities with regulatory support and user value will be key to enabling a secure, efficient, and sustainable energy ecosystem.

Table 7: Abbreviations and acronyms used throughout the paper

| Abbreviation | Definition |
| --- | --- |
| EV | Electric Vehicle |
| V2G | Vehicle-to-Grid |
| V2H | Vehicle-to-Home |
| V2L | Vehicle-to-Load |
| V2V | Vehicle to Vehicle |
| RES | Renewable Energy Resources |
| DER | Distributed Energy Resource |
| HEM | Home Energy Management System |
| APD | Active Power Distribution |
| DR | Demand Response |
| DSO | Distribution System Operation |
| DSR | Demand Side Response |
| DNO | Distribution Network Operation |
| ICT | Information and Communication Technologies |
| IoT | Internet of Things |
| CPS | Cyber-Physical Security |
| MOTEEO | Multi-Objective-Techno-Economic-Environmental Optimization |
| ML | Machine Learning |
| DL | Deep Learning |
| AI | Artificial Intelligence |
| DE | Differential Evolution |
| EVCS | Electric Vehicle Charging Stations |
| VPP | Virtual Power Plant |
| V2B | Vehicle-to-Building |
| IVI | In-Vehicle Infotainment |
| VSI | Voltage Stability Index |

Table 7: Abbreviations and acronyms used throughout the paper

| Abbreviation | Definition |
| --- | --- |
| DSR | Demand Side Response |
| SVR | Support Vector Regression |
| DA | Dragonfly Algorithm |
| BESS | Battery Energy Storage System |
| FCEV | Fuel-Cell Electric Vehicle |
| EVCS | Electric Vehicle Charging Station |
| GA | Genetic Algorithms |
| PSO | Improved Particle Swarm Optimization |
| IDS | Intrusion Detection Systems |
| ZKP | Zero-Knowledge Proofs |
| DTR | Dynamic Thermal Rating |
| ECC | Elliptic Curve Cryptography |
| SiC | Silicon Carbide |
| GaN | Gallium Nitride |
| PUF | Physical Unclonable Function |
| P2P | Peer to Peer |
| PKI | Public Key Infrastructure |
| LSTM | Long Short-Term Memory |
| WOA | Whale Optimization Algorithm |
| GWO | Grey Wolf Optimizer |

## Acknowledgments

None

## Author Declarations

### *Conflict of Interest*

The authors declare that they have no known competing financial interests or personal relationships that could have appeared to influence the work reported in this paper.

### *Author Contributions*

**Bilal Ahmad:** Conceptualization, Methodology, Software, Investigation, Writing – Original Draft.

**Jianguo Ding:** Supervision, Methodology, Project administration, Writing – Review & Editing.

**Tayyab Ali:** Data Curation, Formal analysis, Visualization, Writing – Review & Editing.

**Doreen Sebastian Sarwatt:** Investigation, Writing – Review & Editing.

**Ramsha Arshad:** Investigation, Writing – Review & Editing.

**Adamu Gaston Philipo:** Investigation, Writing – Review & Editing.

**Huansheng Ning:** Supervision, Project administration, Writing – Review & Editing.

**Data Availability**

The data supporting this review are from previously reported studies, which have been cited. No new data were generated.